\documentclass[%
 aip,
 amsmath,amssymb,
 reprint,%
]{revtex4-1}

\usepackage{graphicx}
\usepackage{dcolumn}
\usepackage{bm}

\usepackage[utf8]{inputenc}
\usepackage[T1]{fontenc}
\usepackage{mathptmx}
\usepackage{etoolbox}
\usepackage{xcolor}
\usepackage{enumitem}
\makeatletter
\def\@email#1#2{%
 \endgroup
 \patchcmd{\titleblock@produce}
  {\frontmatter@RRAPformat}
  {\frontmatter@RRAPformat{\produce@RRAP{*#1\href{mailto:#2}{#2}}}\frontmatter@RRAPformat}
  {}{}
}%
\makeatother
\begin{document}

\preprint{AIP/123-QED}

\title{Resonant multi-gap superconductivity at room temperature near a Lifshitz topological transition in sulfur hydrides}
\author{Maria Vittoria Mazziotti}
\affiliation{RICMASS Rome International Center for Materials Science, Superstripes Via dei Sabelli 119A, 00185 Roma, Italy}
\affiliation{Department of Mathematics and Physics, University Roma Tre, via della Vasca Navale 84, 00146 Roma, Italy}
 \author{Roberto Raimondi}
\affiliation{Department of Mathematics and Physics, University Roma Tre, via della Vasca Navale 84, 00146 Roma, Italy}
\author{Antonio Valletta}
\affiliation{Italian National Research Council CNR, Institute for Microelectronics and Microsystems IMM,
	via del Fosso del Cavaliere, 100, 00133 Roma, Italy}
\author{Gaetano Campi}
\affiliation{Institute of Crystallography, CNR, via Salaria Km 29.300, I-00015 Monterotondo, Roma, Italy}
\author{Antonio Bianconi}
\affiliation{RICMASS Rome International Center for Materials Science, Superstripes Via dei Sabelli 119A, 00185 Roma, Italy}
\affiliation{Institute of Crystallography, CNR, via Salaria Km 29.300, I-00015 Monterotondo, Roma, Italy}
\affiliation{National Research Nuclear University MEPhI (Moscow Engineering Physics Institute), 115409 Moscow, Russia}%

\begin{abstract}
The maximum critical temperature for superconductivity in pressurized hydrides appears at the top of superconducting domes in $T_c$ versus pressure curves at a particular pressure, which is not predicted by standard superconductivity theories. The a high-order anisotropic van Hove singularity near the Fermi level observed in band structure calculations of pressurized sulfur hydride, typical of a supermetal, has been associated with the array of metallic hydrogen wires modules forming a nanoscale heterostructure at atomic limit called superstripes phase. Here we propose that pressurized sulfur hydrides behave as a heterostructure made of a nanoscale superlattice of interacting quantum wires with a multicomponent electronic structure. We present first-principles quantum calculation of a universal superconducting dome where $T_c$ amplification in multi-gap superconductivity is driven by the Fano-Feshbach resonance due to configuration interaction between open and closed pairing channels, i.e., between multiple gaps in the BCS regime, resonating with a single gap in the BCS-BEC crossover regime. In the proposed three dimensional (3D) phase diagram the critical temperature shows a superconducting dome where $T_c$ is a function of two variables (i) the Lifshitz parameter ($\eta$) measuring the separation of the chemical potential from the Lifshitz transition normalized by the inter-wires coupling and (ii) the effective electron phonon coupling (g) in the appearing new Fermi surface including phonon softening. The results will be of help for material design of room temperature superconductors at ambient pressure.
\end{abstract}

\maketitle

\begin{quotation}
\end{quotation}

\section{\label{sec:level1}Introduction}

{\color{black}
\subsection{Phenomenological overview}
Pressurized sulfur hydride $H_3S$, with $T_c$=$203 \ K$ at $162 \ GPa$ [\onlinecite{drozdov2015conventional}], has
reached in 2015 the record for the highest critical temperature, held before by cuprate perovskites since 1986
[\onlinecite{bednorz1988perovskite},\onlinecite{gao1994superconductivity},\onlinecite{yamamoto2015high}]. 
This discovery has been
followed by superconductivity in lanthanum hydrides with $T_c$ above $260 \ K$
[\onlinecite{somayazulu2019evidence},\onlinecite{drozdov2019superconductivity}],
 in yttrium hydrides with $T_c= 243 \ K$ 
[\onlinecite{troyan2021anomalous},\onlinecite{kong2021superconductivity}], 
and in a ternary carbonaceous sulfur hydride $CSH_x$
[\onlinecite{snider2020room}] reaching room temperature.
X-ray diffraction, using focused synchrotron radiation, has shown the crystalline $Im\bar{3}m$ lattice symmetry of $H_3S$ above $103\ GPa$ [\onlinecite{einaga2016crystal,goncharov2017stable,duan2017structure,kruglov2017refined,duan2017structure}] 
and X-ray absorption spectroscopy has provided information on the local structure of yttrium hydrides
[\onlinecite{purans2021local}].

Recent experimental results show an anomalous superconductivity phase [\onlinecite{troyan2021anomalous},\onlinecite{kong2021superconductivity}],
while conventional superconductivity
[\onlinecite{eliashberg1960interactions},\onlinecite{dynes1972mcmillan}], 
considering only the pairing of the superconducting electrons via electron-phonon coupling (Cooper pairs) and a single-gap superconductivity paradigm,
has been used to predict and to explain the high critical temperature in pressurized hydrides since the early days [\onlinecite{duan2014pressure,duan2015pressure,durajski2017first,gor2018colloquium,kostrzewa2020lah}].
 
The old single-gap paradigm was found to be incompatible with band structure calculations of $H_3S$ in the pressure range
where the critical temperature shows its maximum value, $T_{c \ max}$.
In fact band-structure calculations [\onlinecite{bianconi2015lifshitz, bianconi2015superconductivity, jarlborg2016breakdown}], 
show that:
\begin{enumerate}[label=(\roman*)]
 \item the applied pressure induces an increasing compressive lattice strain which pushes 
an incipient density of states (DOS) peak, due to a van Hove singularity (vHS), to higher energy until it crosses the Fermi level, 
[\onlinecite{bianconi2015lifshitz}]  as confirmed  by several authors 
[\onlinecite{quan2016van},\onlinecite{souza2017possible}];
\item multiple Fermi surfaces coexist in different  spots of the $k$-space,
 [\onlinecite{bianconi2015superconductivity}];
\item  the Migdal approximation  ${E_{Fn}}$ >> ${\hbar \omega_0}$ in the appearing nth Fermi-surface spot breaks down near the Lifshitz transition
 [\onlinecite{jarlborg2016breakdown}];
\item  the anomalous pressure-dependent isotope coefficient 
[\onlinecite{jarlborg2016breakdown}] strongly deviates from the single-band constant value predicted by the standard BCS theory.
\end{enumerate}
The Fermi energy ${E_{Fn}}$ in the appearing new nth Fermi-surface spot at the Lifshitz transition and the energy width of the vHS singularity
are of the order of  the energy of the optical phonon ${\hbar \omega_0}$=$160\ meV$,  
observed by Capitani et al.[\onlinecite{capitani2017spectroscopic}] in pressurized $H_3S$ infrared spectra. The latter   show a Fano lineshape 
with the characteristic strong asymmetry indicating its interference with electronic degrees of freedom at the Fermi level
[\onlinecite{fano1935sullo},\onlinecite{fano1961effects}].

The vHS  at the Fermi energy, by using band-structure calculations,  has been attributed
to an electronic band of {\it s} orbitals originating from  the network of hydrogen chains with short $H-H$ hydrogen bonds [\onlinecite{jarlborg2016breakdown}]. 
The lattice compressive strain, due to increasing pressure, induces the energy shift of the vHS. The latter  crosses the chemical potential 
yielding a Lifshitz transition for the appearing of a new small Fermi-surface spot [\onlinecite{bianconi2015lifshitz}],
while the other Fermi surfaces contribute to the featureless weak broad background of the density of states.
The Lifshitz transition belongs to the class of electronic topological transitions 
[\onlinecite{lifshitz1960anomalies,volovik2017topological,volovik2017lifshitz,volovik2018exotic}]
for the appearing of a new  Fermi surfaceof the  2.5 order for standard Fermi gases.  These transitions become first order showing arrested phase separation for strongly interacting fermions [\onlinecite{kugel2008model},\onlinecite{bianconi2015intrinsic}].

The critical temperature as a function of pressure in $H_3S$ and $CSH_x$ is shown in Fig.{\ref{fig:1}}.
The external pressure induces a compressive strain, shown in panel (a) of Fig.{\ref{fig:1}}. The strain is given by 
$\epsilon =100 (a -a_0)/a_0$, where
$a_0$ is the lattice constant at $P=103\ GPa$, where the $Im\bar{3}m$  lattice symmetry appears because of a structural phase transition.  
The superconducting $dome$ over the strain range $0.5<\epsilon<4$ shows the  maximum $T_c$ at  $\epsilon=2.5$.
Panel (b) of Fig.{\ref{fig:1}} shows the superconducting $dome$ observed in $CSH_x$ [\onlinecite{snider2020room}]. 
Comparing Panels (a) and (b) we notice that the maximum $T_c$ is higher in the superconducting dome of $CSH_x$ while its width is smaller.

\subsection{The Fano-Feshbach resonance in multi-gap superconductivity}
The paradigm shift to multi-gap superconductivity including the key role of Majorana exchange interaction between different condensates
[\onlinecite{bianconi2003ugo},\onlinecite{vittorini2009majorana},\onlinecite{palumbo2016pion}]
has been proposed since 2015 [\onlinecite{bianconi2015lifshitz}].
The Bogoljubov formulation of superconductivity, beyond the attractive BCS force between two electrons via the exchange of a phonon, includes also the attractive Majorana or repulsive Heisenberg exchange interactions [\onlinecite{bogoljubov1958newmethod}] 
as in nuclear matter. In the latter the forces which are commonly assumed in the phenomenological proton-neutron Hamiltonian include
\begin{enumerate}[label =(\roman*)]
\item the Heisenberg exchange operator for particles which exhibit antisymmetric states, which interchanges both position and spin coordinates [\onlinecite{heisenberg1933structure}];
\item the Majorana exchange operator for particles which exhibit symmetric states, which interchanges the positions of the particles, leaving their spin directions unaffected 
[\onlinecite{majorana1933uber}];
\item the nuclear force, resulting from the exchange of mesons between neighboring nucleons
(Yukawa type) [\onlinecite{yukawa1935interaction}].
\end{enumerate}
The nuclear force has been called also the short range Wigner force, applied exclusively to non-exchange forces to account for the explanation of the large binding energy of He$^4$ in comparison with the deuteron, as well as of the main features of neutron-proton scattering. 
The interplay of these forces in the many-body quantum physical description of nuclear matter has been the object of extended studies
[\onlinecite{iachello2013interacting,feshbach1974interacting,blatt1979nuclear}].

The proposed scenario of multi-gap superconductivity including exchange interactions near a Lifshitz transition in pressurized $H_3S$ [\onlinecite{bianconi2015superconductivity}] 
was previously proposed by Bianconi, Perali and Valletta (BPV) for other non BCS superconductors like  hole-doped cuprates 
[\onlinecite{perali1996gap,valletta1997electronic,bianconi1998superconductivity,bianconi2006multiband,perali2012anomalous}],
 diborides 
[\onlinecite{bianconi2001quantum,bianconi2003ugo,perali2004bcs,perali2004quantitative,bianconi2005feshbach}],
 iron-doped superconductors 
[\onlinecite{innocenti2010resonant,innocenti2010shape,bianconi2012superconductor,innocenti2013isotope,bianconi2013shape}],
 organics [\onlinecite{mazziotti2017possible,pinto2020potassium}],
 metallic nanoscale multilayers with nodal lines where the spin-orbit interaction plays a key role in the $T_c$ amplification [\onlinecite{PhysRevB.103.024523}],
 superconductivity [\onlinecite{bianconi2014shape}] at the interface of oxide perovskites which can host also Majorana fermions  [\onlinecite{mazziotti2018majorana}].
The BPV theory focuses on the quantum Fano-Feshbach resonance 
due to the configuration interaction between the open and the closed scattering channels
[\onlinecite{bianconi2003ugo},\onlinecite{vittorini2009majorana},\onlinecite{palumbo2016pion}]. 
The Fano-Feshbach resonance was first proposed theoretically in atomic physics by Fano in 1935 [\onlinecite{fano1935sullo},\onlinecite{fano1961effects}] 
and extended by Feshbach in 1962 in the many-body physics of nuclear matter [\onlinecite{feshbach1962unified}], where it is called shape resonance and  it is described by the multi-channel optical model. 

\begin{figure}
	\centering
	\includegraphics[scale=0.45]{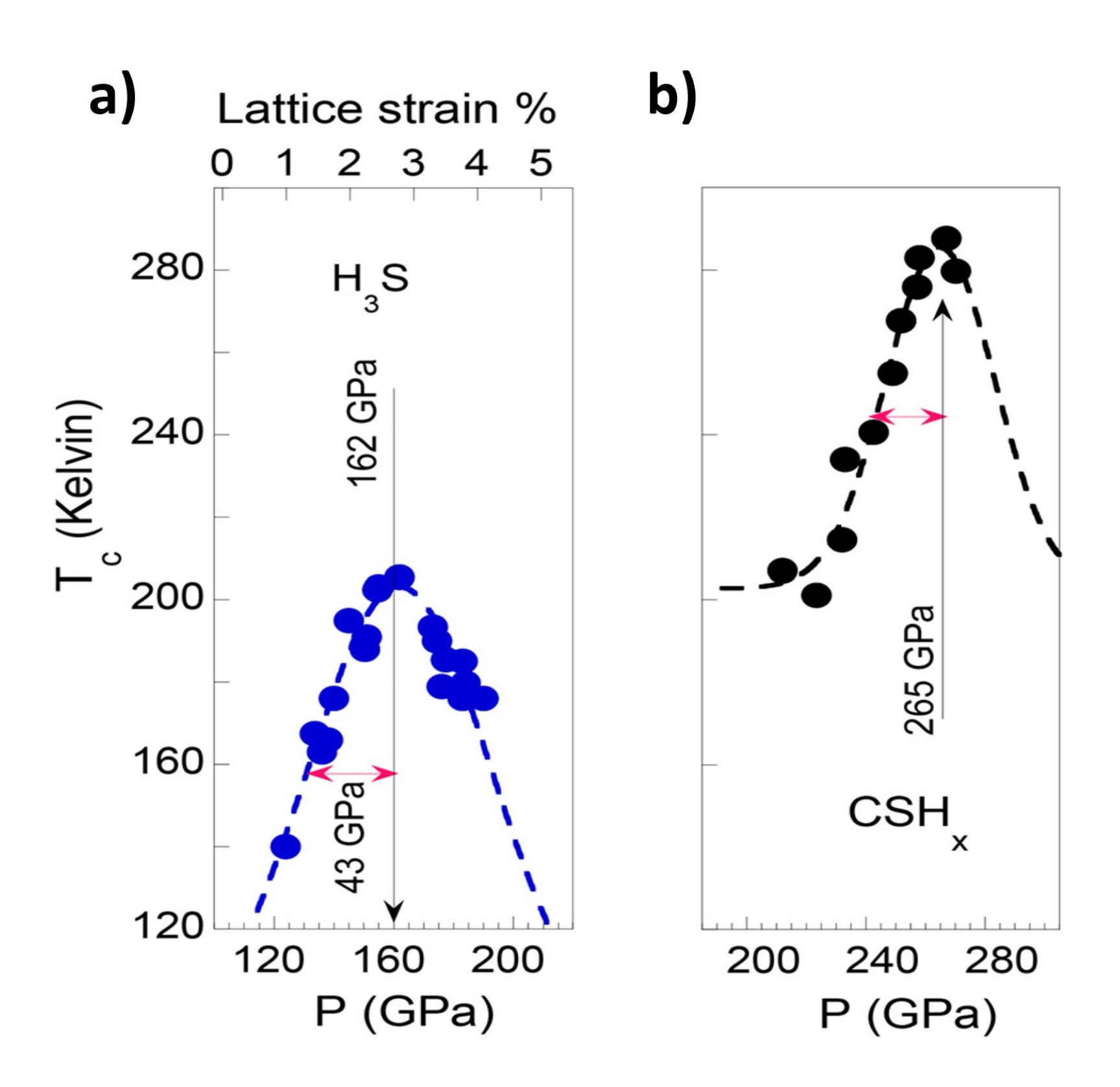}
	\caption{{Panel} \textbf{a}: The superconducting $dome$ of $H_3S$ for $P>120\ GPa$ with the critical temperature $T_{c \ max}=203 K$ at its top
		at $P_{opt}=162\ GPa$ with half width  of about $43\ GPa$.
		{Panel} \textbf{b}:  The superconducting $dome$ of $CSH_x$ for $P>220\ GPa$ with $T_{c \ max}=287 K$
		at 
		$P_{opt}=265\ GPa$}  
	\label{fig:1}     
\end{figure}

In the quantum theory of the many-body systems, made of different electronic components, the Fano-Feshbach resonance appears when the Fermi wavelength of one  of the components is of the order of the system size as in nuclear matter [\onlinecite{feshbach1983conference},\onlinecite{feshbach2014nuclear}] and in condensed matter at the nanoscale [\onlinecite{miroshnichenko2010fano}]. 
Shape resonances have been found in the final states of X-ray absorption near edge structure where the photoelectron wavelength is of the order of interatomic distance 
and the electronic multiple scattering resonance is degenerate with the continuum
[\onlinecite{bianconi1978k,bianconi1980surface,bianconi1982multiple}].

The exchange interaction between condensates was included in the theories for overlapping bands by Suhl, Matthias, and  Walker (SMW)
[\onlinecite{suhl1959bardeen}], Moskalenko [\onlinecite{moskalenko1959superconductivity}] and Kondo [\onlinecite{kondo1963superconductivity}], even though
 they assumed, in a first approximation, that all intraband pairing channels in each of the $n$ bands were in the BCS regime with ($E_{Fn}\gg\omega_0$). Furthermore 
the exchange term for interband pair transfer was assumed to be a constant  parameter with no energy or momentum dependence.  Therefore the above theories of overlapping bands could not include Fano-Feshbach resonances.
Indeed, the Fano-Feshbach resonance in  the Bogoliubov superconductivity theory of multi-gap superconductors is due to the configuration interaction between different pairing channels in different Fermi surfaces  [\onlinecite{mazziotti2021room}] with exchange of pairs between the first condensate in the BCS regime and second condensate in the BCS-BEC crossover regime. In the BCS-BEC regime  the momentum and energy dependence of the exchange interaction between different coexisting gaps plays a key role, while it is neglected in the anisotropic Eliashberg theory of multi-gap superconductors. On the contrary, in the BPV theory [\onlinecite{perali1996gap}], the Fano-Feshbach resonance between a $first$ pairing channel (called closed channel) in the BCS-BEC crossover regime, and the open pairing channels (called open channels) in other large Fermi surfaces in the BCS regime has been calculated from the overlap of the wave-functions of the electron pairs in different bands. The latter are determined by the subtle overlap of the wave-functions of pairs in superlattices of interacting 1D or 2D units. In ultracold fermion gases the Fano-Feshbach resonance has been applied in 2004 to generate unconventional fermion superfluids with a very large ratio of  $T_c/T_F$   [\onlinecite{zwierlein2004condensation},\onlinecite{zhang2004p}].
The quantum amplification mechanism at Fano-Feshbach resonance near a Lifshitz transition
is generated by the  quantum interference of pairing between:
\begin{enumerate}[label=(\roman*)]
\item electrons in  the new appearing small Fermi surface with low Fermi energy 
and Fermi wavelength  $\lambda_F$ of the order of the system size; 
\item electrons in other Fermi surfaces with very high Fermi energies and very short Fermi wavelength $\lambda_F$. 
\end{enumerate}
At optimum $T_c$ in the closed pairing channel,  $\lambda_F$ is larger but 
close to the superconducting coherence length $k_F\xi_0 \sim 10$ [\onlinecite{uemura1989universal},\onlinecite{pistolesi1994evolution}].

\subsection{Choice of the model}
Here we propose that the experimental superconducting $dome$, given by the curves of the critical temperature $T_c$ as a function 
of pressure P in Fig.{\ref{fig:1}} in sulfur hydrides,  is the smoking gun of the Fano-Feshbach resonance between pairing channels driven by the variable lattice strain [\onlinecite{perali1996gap},\onlinecite{mazziotti2017possible},\onlinecite{agrestini2003strain}] which tunes the chemical
potential at a topological Lifshitz transition. 

The Fano-Feshbach resonance, in multi-gap superconductivity in $H_3S$, is supported by the unusual behavior of the isotope coefficient. Indeed, the isotope coefficient decreases from $2.37$ to $0.31$ in the range going from the threshold to the top of the superconducting
$dome$ [\onlinecite{jarlborg2016breakdown,szczkesniak2017isotope,szczkesniak2018unusual,drozdov2019superconductivity}],
deviating markedly from the value $ 0.5$, predicted by the single-band BCS theory. A similar anomalous behavior of the isotope 
coefficient has been found in the superconducting $dome$ of cuprate perovskites 
[\onlinecite{perali1997isotope}],[\onlinecite{bianconi1998superconductivity}],[\onlinecite{perali2012anomalous},\onlinecite{innocenti2013isotope}].

\indent A nanoscale heterostructure is expected to appear in compounds made by combining several chemical elements  which leads to competing  orders of  electronic degrees of freedom.
In pressurized hydrides the nanoscale heterogeneity is determined by the  local lattice structure controlled by the effects of the lattice strain resulting from the interplay between lattice misfit-strain (or chemical pre-compression) and the external pressure. 
These heterostructures are a particular case of a supersolid stripes crystal [\onlinecite{bianconi2013superstripes}] called superstripes  [\onlinecite{bianconi2013shape,bianconi2013shapeb}],
which can be realized with optical lattices in ultracold  gases [\onlinecite{masella2019supersolid}].

\begin{figure}
	\includegraphics[scale=0.7]{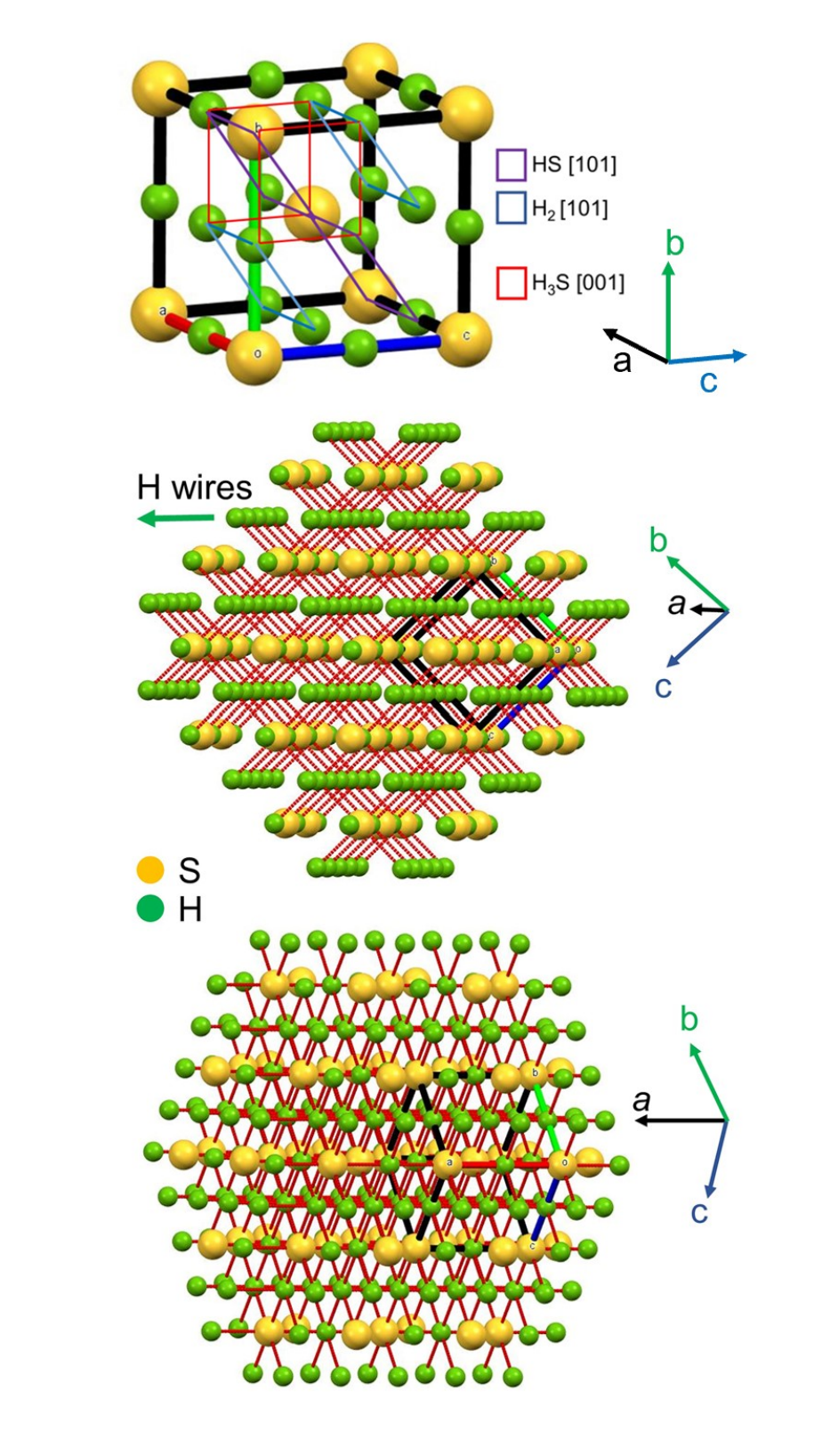}
	\caption{The upper panel shows the unit cell of pressurized $H_3S$ crystal structure at $150\  G Pa$ with cubic $Im\bar{3}m$ lattice.  
		The central and lower panels show that at mesoscale $H_3S$  appears to be made of stacks of 
		$H_2$ layers (small green dots) in the [101] plane  intercalated by $HS$ layers where $S$ is indicated by large yellow dots. 
		The central projection of the $H_3S$ crystal shows that the
		$H_2$ layers are made of atomic hydrogen chains with the shortest H-H metallic bonds in the $[100]$ direction.
	}
	\label{fig:1b}     
\end{figure}

The superconductivity in the superstripes phase with coexisting different localised and delocalised electronic components moving in complex nanoscale heterostructures of low dimensional  (quasi one-dimensional 1D) structural units (called chains or stripes or ladders) 
has been found in:
\begin{enumerate}[label=(\roman*)]
\item A15 intermetallics which have held  the record for the highest superconducting  $T_c=23.2\ K$ from 1973 to 1986. A15 intermetallics like $Nb_3Ge$ and $Nb_3Sn$ have the same average crystal symmetry $Im\bar{3}m$ as $H_3S$ [\onlinecite{mazziotti2021room}] and show complex textures made of a metallic 3D network of interacting 1D metallic Nb chains [\onlinecite{testardi1975structural}].
\item  hole doped cuprate perovskites where 2D networks of extrinsic stripes with different local lattice distortions 
[\onlinecite{bianconi1991linearly,bianconi1996determination,bianconi1996stripe}] 
appear at nanoscale in the $CuO_2$ atomic layers
facilitated by the polymorphism of perovskite structures. These
 form metamorphic lattice stripes in mismatched material systems
[\onlinecite{gavrichkov2019there}] whose mismatch is tuned by the lattice misfit strain [\onlinecite{agrestini2003strain},\onlinecite{bianconi2000strain},\onlinecite{di2000evidence}].
\item  superconducting organics like doped p-terphenyl [\onlinecite{pinto2020potassium}], where 1D-wires of short  hydrogen bonds  have been observed by X-ray diffraction [\onlinecite{barba2018anisotropic}] and it was proposed that the high-$T_c$ is driven by the Fano-Feshbach resonance in the nanoscale superlattice of quantum wires [\onlinecite{mazziotti2017possible}].
\end{enumerate}

Superlattices of two-dimensional (2D) metallic quantum wells at nanoscale have been found in: 
\begin{enumerate}[label=(\roman*)]
\item diborides made of stacks of boron layers intercalated by magnesium [\onlinecite{bauer2001thermal,agrestini2001high,bianconi2001superconductor,agrestini2004substitution,di2002amplification}];
\item  iron-based perovskite superconductors, iso-structural with electron doped cuprates, [\onlinecite{ricci2009microstrain},\onlinecite{ricci2010structural}], 
which are made of stacks of iron atomic layers and the tuning of the chemical potential near the Lifshitz transition has been clearly seen in ARPES experiments 
[\onlinecite{bianconi2013shape},\onlinecite{kordyuk2018electronic},\onlinecite{pustovit2016metamorphoses}].
\end{enumerate}

 In this work we present the theoretical prediction of the superconducting dome for room-temperature superconductivity in pressurized hydrides 
due to a Fano-Feshbach resonance  near a Lifshitz transition
 in the frame of the multi-gap superconductivity scenario discussed recently by several authors
[\onlinecite{
guidini2016bcs,
cariglia2016shape,
doria2016multigap,
bussmann2016high,
valentinis2016bcs,
bussmann2017road,
bussmann2019multi,
chubukov2018evolution,
salasnich2019screening,
kagan2019fermi,
tajima2020investigate,
vargas2020crossband}]. 

A key feature of our approach is the inclusion of the exchange integrals, obtained by the overlap of the wave-functions of electrons in different Fermi surfaces.  
We evaluate them by solving the Schr\"odinger equation for a lattice heterostructure including the renormalisation of the chemical potential at the Lifshitz transition
 with the opening of a new superconducting gap, controlled by the constraint of the  number density equation. 
The results provide a significant step in understanding room-temperature  superconductivity and the physical origin of the superconducting \textit{dome}.  
Moreover, the results indicate  a road map for the material design of artificial mesoscopic heterostructures made of nanoscale quantum wires which  can be used by 
material scientists to synthesize new  room-temperature superconductors at ambient pressure.

}

\section{The van Hove Singularity in the superstripes phase}
\begin{figure}
	\includegraphics[scale=0.8]{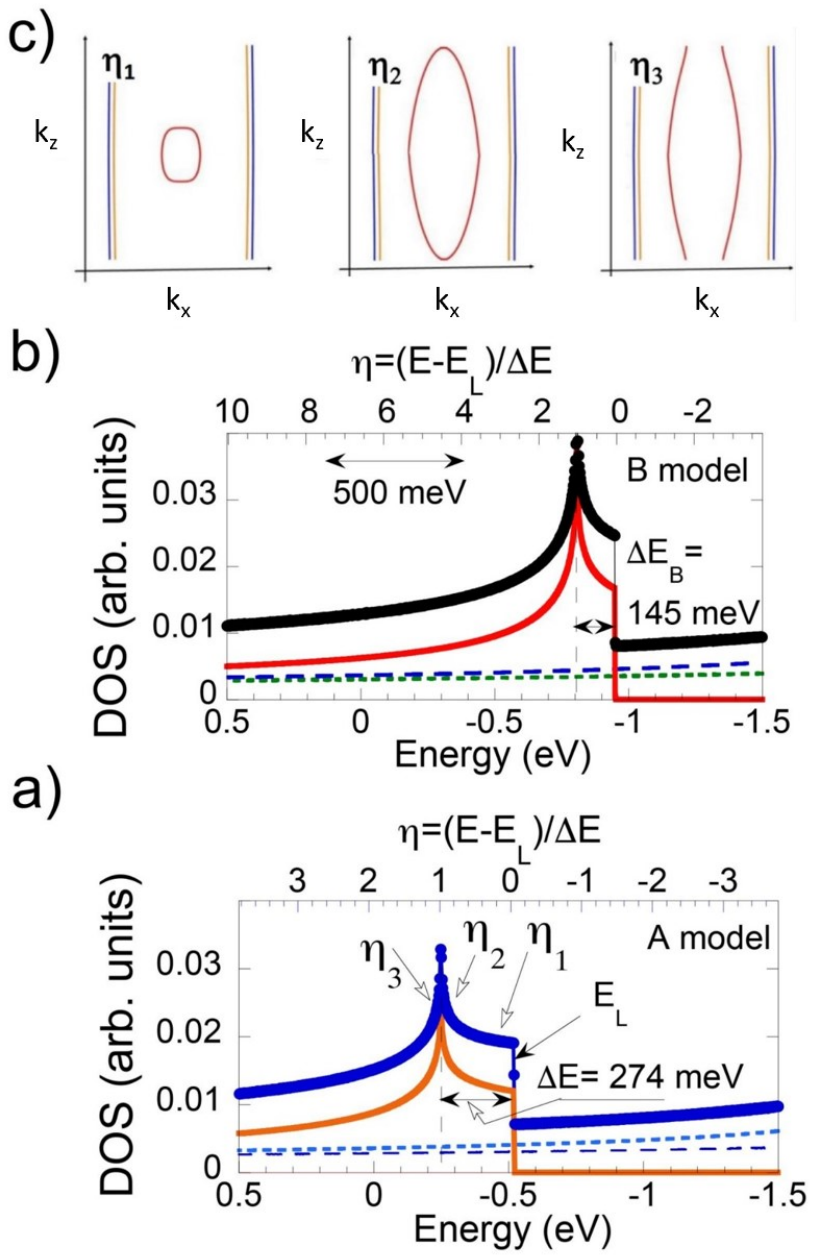}
	\caption{\textit{Bottom panels}: total Density of States (DOS)  for the case A, panel  \textbf{a} (for  the case  B, panel \textbf{b}) shown by the thick solid blue (black) line as a function of the Lifshitz parameter $\eta$ of the superlattice of quantum wires with weak inter-wire interaction giving the transversal dispersion $\Delta E = 274\ meV$ ($145\ meV$). The figure shows also the high partial DOS curves at the van Hove singularity due to the upper third subband (red line)  with small Fermi energy and the low partial DOS curves due to first and second  subbands (blue dashed lines) with  high Fermi energies. 
	The top panel \textbf{c} shows the Lifshitz topological transition in the Fermi surfaces due to the third subband, shown with a solid red line, where the Fermi surfaces for the first and second subbands are indicated by blue and orange lines. The appearing Fermi surface due to the third subband changes at three different values of $\eta$ indicated in  panel  \textbf{a}: $ \eta_1 $  corresponds to the appearing of a small tubular Fermi surface in the $k_x, k_z$ plane;  $\eta_2 $ is the energy where the size of the tubular Fermi surface becomes large and it is close \textcolor{red}{to} the Lifshitz transition for neck disrupting (or opening a neck) where its topology  changes from 2D for $\eta_2 $  to 1D for $\eta_3$ [\onlinecite{bianconi2015lifshitz}].
	}
	\label{fig:2}     
\end{figure}

\begin{figure}
	\includegraphics[scale=0.7]{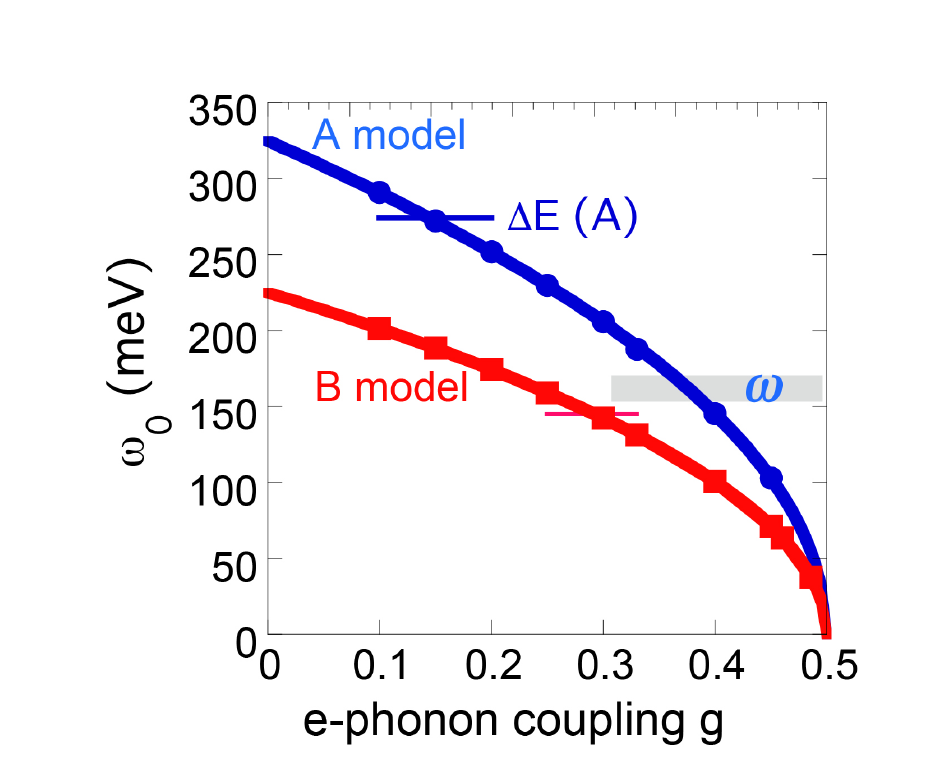}
	\caption{The softening of the phonon energy $\tilde{\omega}_0$ according to the  Migdal theory, as a function of the intraband electron-phonon coupling $g$  in the small Fermi surface spot appearing at the Lifshitz transition for the  case A (blue line)  and for the  case B (red line)}
	\label{fig:3}     
\end{figure}

{\color{black}We propose that  in agreement with  [\onlinecite{bianconi1996high},\onlinecite{bianconi2001process},\onlinecite{bianconi1994possibility}] a heterostructure at atomic limit made of superconducting quantum wires running in the $x$-direction intercalated by spacers of thickness $W$ with periodicity $d = W + L $ (in the $z$-direction). The metallic wires are separated by spacers which generate a potential barrier  of amplitude V  [\onlinecite{bianconi1996high,bianconi2001process,bianconi1994possibility}], where
the parameters values  are chosen in order to capture the main features of the DOS at the van Hove singularity near the Fermi level of $H_3S$.}
The lineshape of the DOS peak shows the features of a high-order anisotropic van Hove singularity typical of a supermetal [\onlinecite{isobe2019supermetal}],  In this superstripes phase the superconductivity is calculated using the BPV approach first proposed for striped cuprate perovskites
[\onlinecite{perali1996gap, 
perali1997isotope,
bianconi1997high,
valletta1997electronic,
bianconi1998superconductivity,
bianconi2006multiband,
perali2012anomalous}].

We present numerical calculations of the multi-gap superconductivity $dome$ of the critical temperature as a function of  pressure where we change both 
$i)$ the proximity of the chemical potential \textcolor{black}{to} the Lifshitz transition,  
$ii)$ the electron-phonon coupling for electrons in the upper subband
and the $iii)$ the phonon softening for increasing electron-phonon coupling g in the disappearing Fermi surface.

The proximity to the Lifshitz transition is measured by the Lifshitz parameter $\eta$  
given by the energy difference between the chemical potential and the energy $E_L$ of the topological Lifshitz transition, which is the band-edge energy of the highest energy subband, normalized to the transversal energy dispersion, $\Delta E$, between the 1D metallic chains
$$
\eta=\frac{\mu-E_L}{\Delta E}.
$$

The applied external pressure induces  the variation  of either $i)$  of the Lifshitz parameter $\eta$ and $ii)$ of the electron-phonon coupling  joint with phonon energy softening of the particular phonon mode coupled with the electrons in the small Fermi surface spot in the new appearing subband.  In the heterostructure of quantum wires the electrons along the $x$-direction are free, while along the $z$-direction they are subjected to a periodic potential. Hence, the eigenfunctions $\psi_{nk_z}(z)$ and the eigenvalues $E_{n}(k_z)$, along the confinement direction, can be computed only numerically
by solving a corresponding Kronig-Penney model. The solution of the eigenvalues 
equation gives the electronic dispersion for the $n$ subbands.
Indeed, in the heterostructure of quantum wires, the quantum-size effects give a multiband electronic structure where the subband with 
higher energy shows a two-dimensional behavior due to hopping between 1D-chains.
For the numerical calculation of the superconducting dome of $H_3S$ we start with the evaluation of the DOS peak in [\onlinecite{jarlborg2016breakdown}] at $E_F$ by using a model of 1D-chains corresponding to the chains with short H-H bonds, as shown in Fig.{\ref{fig:1b}}.
We have designed an artificial nanoscale heterostructure at atomic limit made of quantum wires of width $L= 0.85$ nm, spacers of width $W= 0.55$ nm separated by a potential barrier V=4.16 eV which reproduces the van Hove singularity in a range of 500 meV around the Fermi energy  $H_3S$ i.e., within the energy cut-off  of the pairing interaction relevant for the emergence of superconductivity. 
We consider the models  A and B for the heterostructure characterized by different coupling between the quantum wires i.e., with different transversal dispersion $\Delta E$ obtained by changing the effective mass in the spacers and in the wires. 
The model A of the superlattice of wires is characterized  by  $\Delta E$= $274\ meV$  transversal dispersion as it was found in $H_3S$.
The model B is characterized by a smaller transversal dispersion $\Delta E=145\ meV$, obtained by increasing the effective mass in the barrier. A smaller dispersion is expected to give a sharper superconducting $dome$ 
and a higher maximum critical temperature close to room temperature.
The DOS peak and the partial DOS for the model A  and model B as a function of the Lifshitz parameter 
are plotted in panel (a) and panel (b) of Fig.{\ref{fig:2}} respectively.
Panel (c) of  Fig.{\ref{fig:2}} shows the Lifshitz transition for the appearing of a new Fermi surface,
called of  type (I), tuning the chemical potential near the band-edge ($\eta_1$) of the subband with 
a critical point  where a new 2D Fermi surface spot appears.
The second type of Lifshitz transition (type II) occurs at the opening of a neck in the small Fermi surface with the appearing of a singular nodal point 
which gives the sharp DOS maximum at $\eta_2$ (Fig.{\ref{fig:2}} panel (c)) at the crossover between the 2D and 1D topology. The nearly flat portion of the DOS between  (type I)   and  (type II)  Lifshitz transitions in Fig.{\ref{fig:2}} correspond with the regime where a small 2D Fermi surface with a low Fermi energy appears.
While in previous theoretical descriptions of the Fano-Feshbach resonance near the Lifshitz transition the electron-phonon coupling $g_{nn}$ was assumed to be constant, in this work we take into account that the external pressure changes either the Lifshitz parameter ($\eta$)  and the electron-phonon coupling $g_{nn}$ in the appearing Fermi surface in the upper subband  and the renormalized phonon energy  $\tilde{\omega}_0$ shows the softening according to the \textit{Migdal relation}:
\begin{equation}
\tilde{\omega}_0=\omega_0  \sqrt{1-\textcolor{black}{2 \times \max_n g_{nn}}}.
\label{Mig}
\end{equation}
The relation contains the coupling constant for the metal forming the superconducting layers for $g<0.5$,
and it is used here to qualitatively estimate 
the effect of the coupling constant on the phonon frequency in the appearing Fermi surface as it is shown in Fig.\ref{fig:3}. In our theory, the  variable $\tilde{\omega}_0$ is also the cut-off energy of the pairing interaction in the Bogoliubov gap equation which changes with $\eta$.
We have fixed, for the case A, $\omega_0=330\ meV$ in order to reproduce with moderate intraband electron phonon coupling, $0.3<g<0.33$, the experimental phonon frequency, $\tilde{\omega}_0=160\ meV$  [\onlinecite{capitani2017spectroscopic}], measured in pressurized $H_3S$ at $150\ GPa$ 
For the case B we have fixed  $\omega_0=225\ meV$, in order to get $\tilde{\omega}_0 = \Delta E$ = $145 \ meV $ with moderate intraband electron phonon coupling $g=0.25$.

In the case of organic superconductors [\onlinecite{mazziotti2017possible}] it has been shown that  the amplification of the critical temperature in heterostructures of quantum wires and a narrow superconducting dome occurs where the coupling in the appearing new nth Fermi surface is larger  than the in other Fermi surfaces and the interband coupling is small. In our model $ g_ {nn'} $ is the superconducting  dimensionless
coupling constant for the three-band system which has a matrix structure that depends on the band indices $n$ and $n'$. 

In this work we confirm previous results  [\onlinecite{mazziotti2017possible}]: in fact  the superconducting dome 
with a sharp drop of $T_c$ at both sides of the maximum
with a stronger Fano-Feshbach anti-resonance is generated by a weak intra-band coupling for the Cooper pairing channel $g_{nn}$ and weak inter-band exchange channels  $g_{nn'}$.
The Fano-Feshbach resonance increases the maximum value  of $T_c$ at the top of the $dome$ increasing the intra-band coupling for the Cooper pairing channel $g_{nn}$ in the new appearing or disappearing Fermi surface. 
Moreover the maximum of the critical temperature is expected to increase where the phonon energy giving the energy cut off for the pairing processes is of the same order as the hopping energy between the wires $\Delta E =\tilde{\omega}$.

When the  Lifshitz parameter is tuned between the band-edge and the van Hove singularity, 
a new Fermi surface appears with a very small number density of electrons in the strong coupling limit. 
The condensates in the other Fermi surfaces (first and second subband)  have a very high Fermi energy and therefore are in the adiabatic regime and coexist with a third condensate in the small Fermi surface where the classical BCS approximations are violated. In the models (A) and (B) for 0<$\eta$<1  a new closed 2D Fermi surface appears as shown in band structure calculations 
[\onlinecite{bianconi2015lifshitz}] for $H_3S$ around 160 GPa where the maximum critical temperature is observed at a top of a $dome$.
For this heterostructure we assume that quantum size effects are not negligible and the electron hopping in the transverse direction is finite so that the quantum wires can be considered to be interacting wires in the metallic phase while very weakly interacting wires are in the localization limit. This is reflected in the spectrum that appears to split into $n$ subbands characterized by quantized values of the transverse moment that depends on the band index and the dimension of the wires.

In the heterostructure of quantum wires the electrons along the $x$-direction are free, while along the $z$ direction they are subjected to a periodic potential $V(z)$:

\begin{equation}
V(z)=V \sum^{\infty}_{-\infty} \theta (W/2-|m\lambda_p-z|).
\end{equation}

In the periodic potential we assume that the full single-particle wave-function can be written as
\begin{equation}
\label{eq:4.1}
\psi_{n, \mathbf{k},\alpha}(\mathbf{r})=\frac{1}{\sqrt{L_xL_z}}e^{ik_xx}\psi_{nk_z}(z)\boldsymbol{\chi}_\alpha,
\end{equation}
where $L_x$ and $L_z$ are the spatial dimensions of the system, $n$ is the band index, $\mathbf{k}=(k_x,k_z)$ is the wavevector, and $\boldsymbol{\chi}_\alpha$ is the spinor part with spin $\alpha=\uparrow$ or $\downarrow$. The corresponding energy eigenvalues, independents from the spin, are given by
\begin{equation}
\label{eq:4.2}
\varepsilon_{n}(\mathbf{k})=\frac{\hbar^2}{2m_x}k^2_x+ E_{n}(k_z).
\end{equation}

The eigenfunctions $\psi_{nk_z}(z)$ and the eigenvalues $E_{n}(k_z)$ are computed numerically
 by solving a corresponding Kronig-Penney model. The solution of the eigenvalues 
equation gives the electronic dispersion for the $n$ subbands.

\section{multi-gap superconductivity beyond BCS}

In the multi-gap superconducting scenario the exchange integral for pairs of electrons in different bands  plays a key role for the $T_c$ amplification while it is neglected in the single-band BCS theory.

{\color{black}The pairing interaction is assumed to be originated from
an electron-electron contact interaction with a cut-off equal to the renormalized phonon energy $\tilde{\omega}_0$ .   
The pairing interaction takes then a generalized BCS  form}
\begin{equation}
\label{eq:4.4}
U_{{\bf k}{\bf k'} }^{nn'}=\tilde{U}_{k_zk'_z}^{nn'}\theta(\tilde{\omega}_0-|\xi_{n,{\bf k}}|)\theta(\tilde{\omega}_0-|\xi_{n',{\bf k'}}|)
\end{equation}
\textbf{where} $\xi_{n,{\bf k}}=\varepsilon_{n}(\mathbf{k})-\mu$.
 {\color{black}
The $\tilde{U}_{k_zk'_z}^{nn'}$ coupling constants, which in the original BCS model are structureless,  originate from the matrix elements between exact eigenstates of the superlattice, and   depend on the wave vectors $k_z$ and $k_z'$  in the superlattice direction as well as on the band indices $n$ and $n'$. This induces a structure in the k-dependent interband coupling interaction for the electrons that determines the quantum interference between electron pairs wave functions 
in different subbands or minibands of the superlattice. 
The generalized couplings can be expressed as}
\begin{equation}
\nonumber
\tilde{U}_{k_zk'_z}^{nn'}=  -\textcolor{black}{U_{0}^{nn'}} \  I_{k_zk'_z}^{nn'}  ,
\end{equation}
{\color{black} where $-\textcolor{black}{U_{0}^{nn'}}$ is the original {\it attractive} contact interaction, 
\textcolor{black}{that we allow to have a dependence from the band index of the electron pairs},
and $I_{k_zk'_z}^{nn'}$
is the pair superposition integral, calculated considering the interference between electronic wave functions in different subbands [\onlinecite{innocenti2010resonant}]. 
}
$$
I_{k_zk'_z}^{nn'}=\frac{1}{L_x L^2_z} \int  \psi_{nk_z}^{*}(z) \psi_{n-k_z}^{*}(z)  \psi_{n'k'_z}(z) \psi_{n'-k'_z}(z) dz.
$$
Notice that for vanishing $V$, the amplitude  of the periodic potential, the overlap integrals would reduce to the standard BCS form
$I_{k_zk'_z}^{nn'} =(L_xL_z)^{-1}$ independent of the $k_z$ and $k_z'$ wave vectors as well.
We emphasize that the  exchange  interaction is not constant but depends not only on the wave vector along z but also on the band index, therefore it has a matrix structure.   For later reference and to compare with the homogeneous case,  it is useful to introduce the standard dimensionless coupling constant \textcolor{black}{$g_{nn'}=U_{0}^{nn'} N_0$}, where $N_0$ is the two-dimensional density of states.
The non diagonal  terms  ($nn'$) of the superposition integral are calculated for model A and are shown in Fig.5.

\begin{figure}
	\includegraphics[scale=0.2]{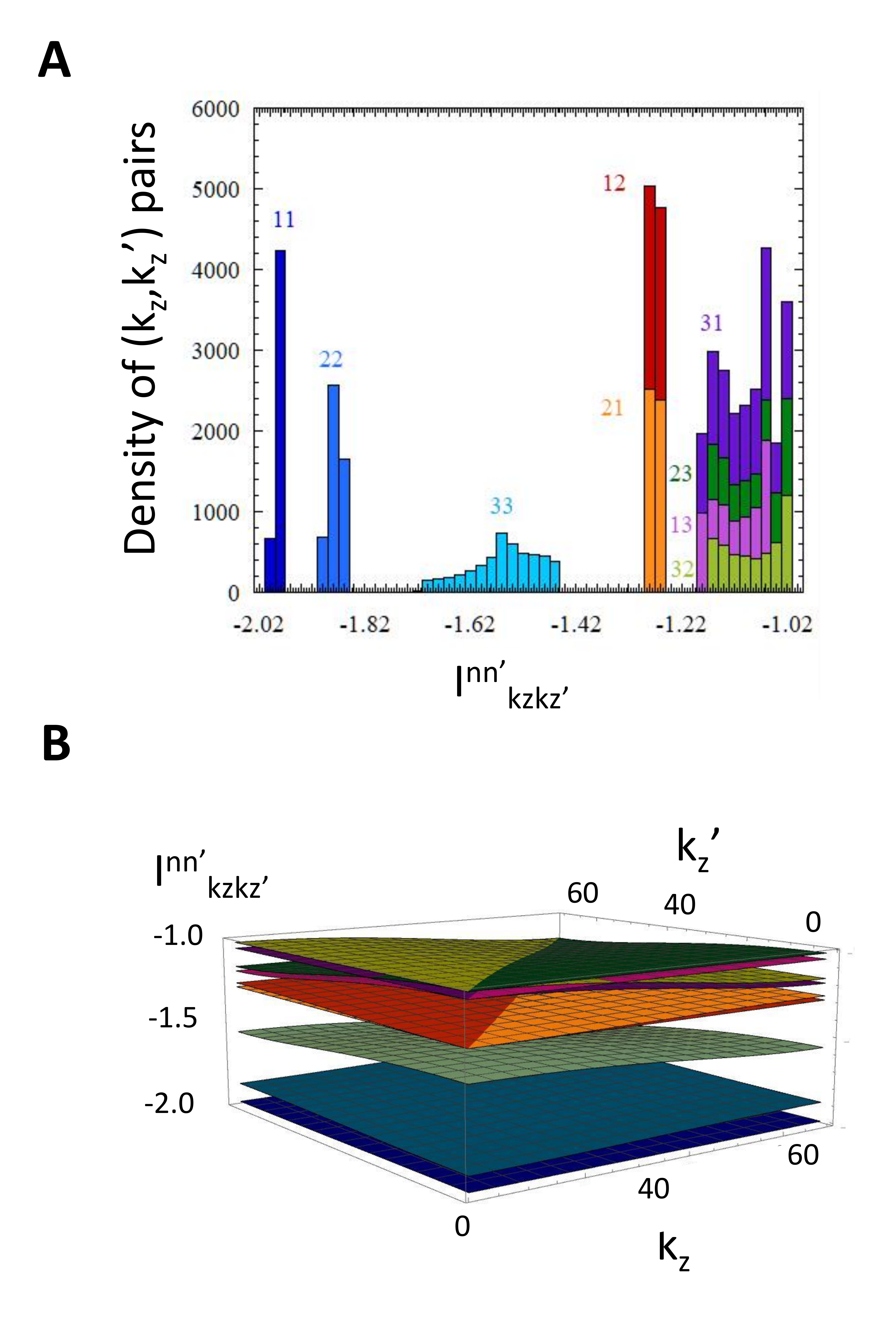}
	\caption{\textit{Panel}  \textbf{A}: histogram of the superposition integral of equation  (\ref{eq:4.4}). The numbers indicate the different elements of the matrix of the intraband pairings $ I_{{k}_z{k}'_z}^{n,n'} $ and the interband couplings $ I_{{k}_z{k}'_z}^{n,n'}$. \textit{Panel} \textbf{B}:  the matrix elements of the exchange integral as a function of the wavevectors in the direction of the confinement potential. The colors correspond to those of the histogram.}
	\label{fig:4.2}     
\end{figure}
The self-consistent equation for the superconducting gap at zero temperature can then be written as
\begin{equation}
\label{eq:4.5}
\Delta_{n k_z} = -\frac{1}{2} \sum_{n',\mathbf{k}'} \frac{U^{nn'}_{{\bf k}{\bf k'}}\Delta_{n'k'_z}}{\sqrt{(\varepsilon_{n'}(\mathbf{k'})-\mu)^2+|\Delta_{n'k'_z}|^2}},
\end{equation}

In order to take into account  the renormalisation of the chemical potential and charge densities in each subband when a new superconducting gap appears in a single subband,  the joint Bogoliubov gap equation and the charge density equation have been solved where the
charge density $\rho$ and the chemical potential in the superconducting phase are related by
\begin{equation}
\label{eq:4.53}
\rho=\frac{1}{L_x L_z}\sum_{n \mathbf{k}} \bigg(1- \frac{\varepsilon_{n}(\mathbf{k})-\mu}{\sqrt{(\varepsilon_{n}(\mathbf{k})-\mu)^2 +\Delta^2_{n,k_z}}}\bigg) \theta(\mu-\varepsilon_{n}(\mathbf{k})),
\end{equation}
The joint solution of the gap equation (3) and the density equation (4) is essential in order to correctly describe the multi-gap superconductivity near the Lifshitz transition where the gap in the  upper subband approaches to the Bose-Einstein condensation. 

The superconducting critical temperature is calculated by iteratively solving the linearized equation
\begin{equation}
\Delta_{n\mathbf{k}}  = -\frac{1}{2} \sum_{n'\mathbf{k}'}  U_{{k}_z{k'}_z}^{nn'}  \Delta_{n'k'_z} \frac{ \tanh \left(\frac{(\varepsilon_{n'}(\mathbf{k}')-\mu)}{2T_c}\right)}{(\varepsilon_{n'}(\mathbf{k}')-\mu)}
\label{eq:4.6}
\end{equation}
until the vanishing solution is reached with increasing temperature.

Here we present a case of Fano-Feshbach resonant superconductivity giving a superconducting $dome$ where the top of the $dome$ reaches the high temperature range $200<T_c \ max<300 K$ of pressurized hydrides much larger than $T_c \ max=123 K$ calculated in a previous work for cuprates and organics  [\onlinecite{mazziotti2017possible}]. This result is obtained by the resonance regime by increasing gaps anisotropy where the two gaps differ by a sizable 
factor in the range 2.9-3.9, at the top of the dome where the coupling strength in a small Fermi surface spot is in the range  0.3<g<0.42
and the phonon energy scale determines not only a large prefactor for the critical temperature, but it also induces a large width of the resonance.

Here, following Ref.[\onlinecite{mazziotti2017possible}], the superconducting dome is generated by considering the case where the first and the second subband are in a weak coupling regime \textcolor{black}{
because the Fermi level is very far from the band edge.
Therefore in our model we 
fixed the values of $U_{0}^{nn'}$ in order to have the following values for the dimensionless coupling constants: $g_{11}=g_{22}=0.10$. On the contrary the coupling in the third subband $g_{33}=g$ is considered to be variable because the Fermi level is tuned around the band edge. In fact we expect that the electron-phonon coupling for the upper subband should be enhanced because of a Kohn anomaly or because the interplay with the formation of a charge density wave (CDW) in a narrow momentum region around the CDW wave vector. In  parallel, 
we vary the cut-off energy $\tilde{\omega}_0$ according
to the Migdal relation [\onlinecite{migdal1958interaction}].}
\begin{figure}
	\includegraphics[scale=0.3]{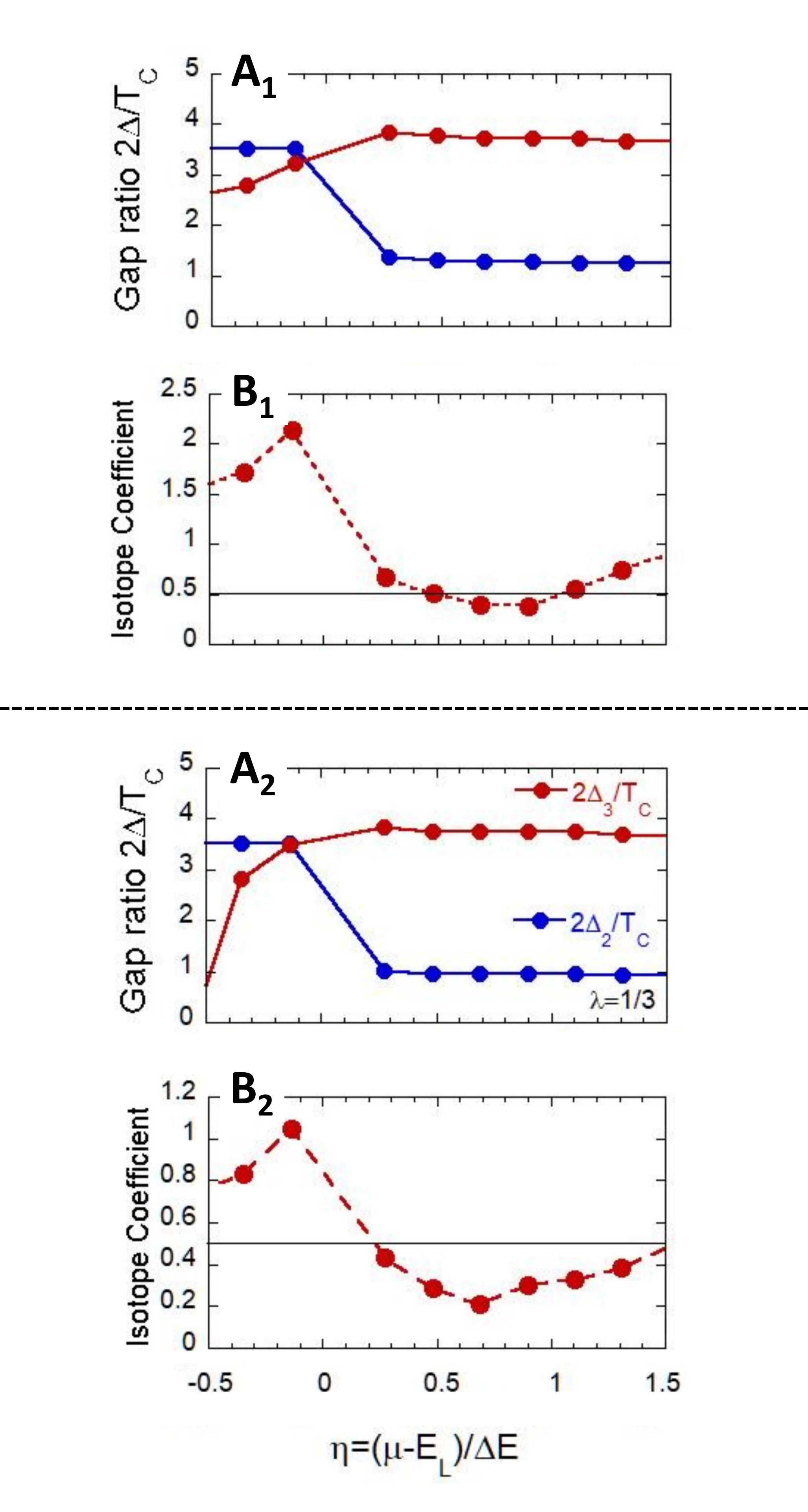}
	\caption{\textit{Panel} $\mathbf{A_1}$: gap ratio as a function of the Lifshitz parameter for $ g =1/4 $  (case (A)) for the second and third subband. \textit{Panel} $\mathbf{B_1}$: isotope coefficient as a function of the Lifshitz parameter for $ g= 1/4 $ (case (A)). \textit{Panel} $\mathbf{A_2}$: gap ratio as a function of the Lifshitz parameter for $ g=1/3 $ (case (B)) for the second and third subband. \textit{Panel} $\mathbf{B_2}$: isotope coefficient as a function of the Lifshitz parameter for $ g = 1/3 $ (case (B)).}
	\label{fig:11}     
\end{figure}

\begin{figure}
	\includegraphics[scale=0.3]{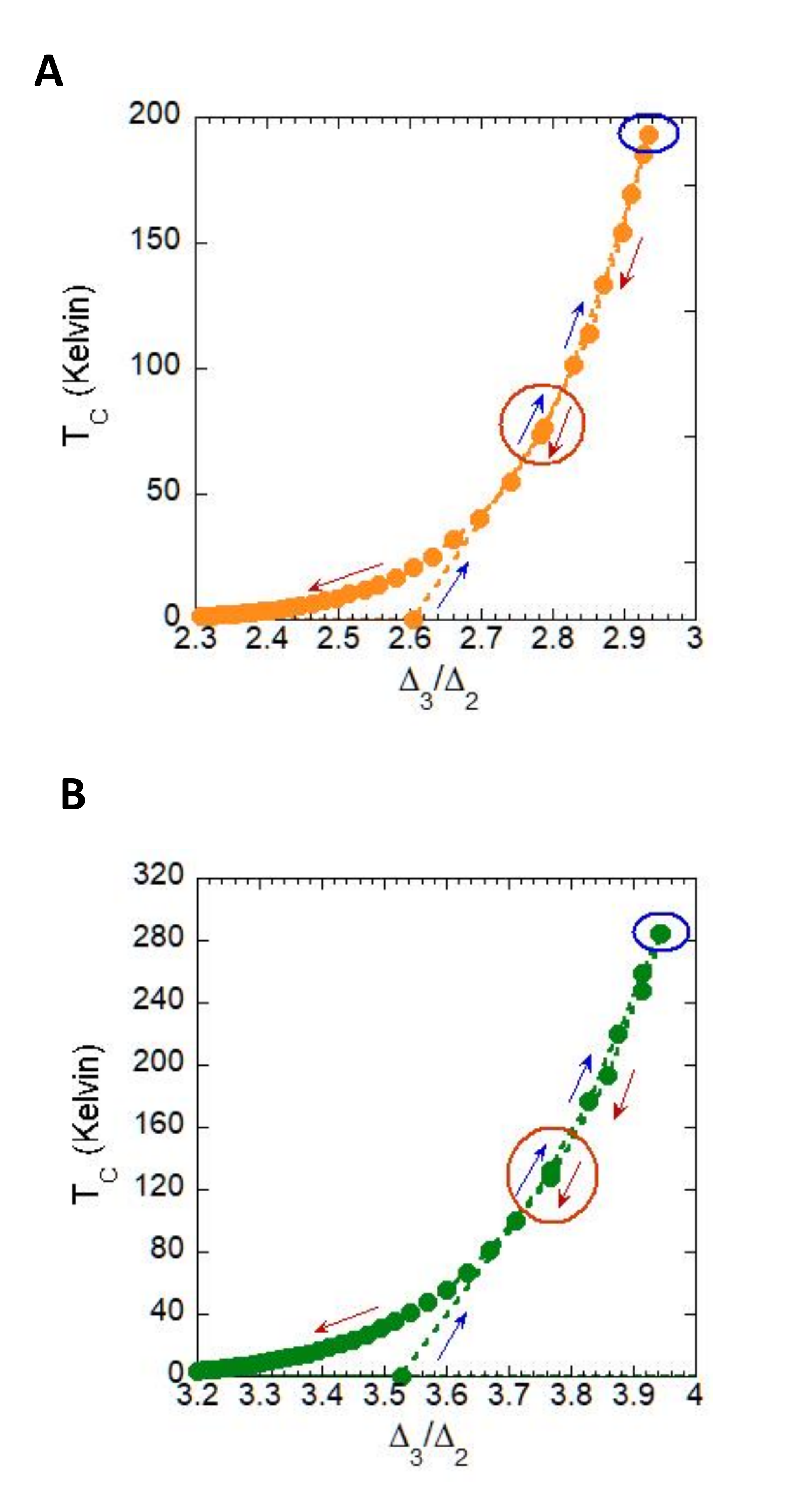}
	\caption{The critical temperature as a function of the ratio between the gap in the third subband and the gap in the second subband. Panel  \textbf{A} represents the trend for the case (A), while panel \textbf{B} represents the case (B). It can be noted that in the range $ 2.6<\Delta_3 / \Delta_2<2.9 $ (for the case (A)) or $ 3.5<\Delta_3 / \Delta_2<3.9 $ (for the case (B))  the critical temperature increases as the anisotropy between the gaps increases (blue arrows) until it reaches a maximum value when $ \Delta_3 / \Delta_2 $ is maximum, from this point on then the $ T_c $ decreases almost exponentially as the ratio between the gaps decreases (red arrow). The blue circle represents the point where $ T_c $ is maximum, the red circle the point of intersection of the two opposite trends.}\label{fig:12}     
\end{figure}

\begin{figure}
	\includegraphics[scale=1]{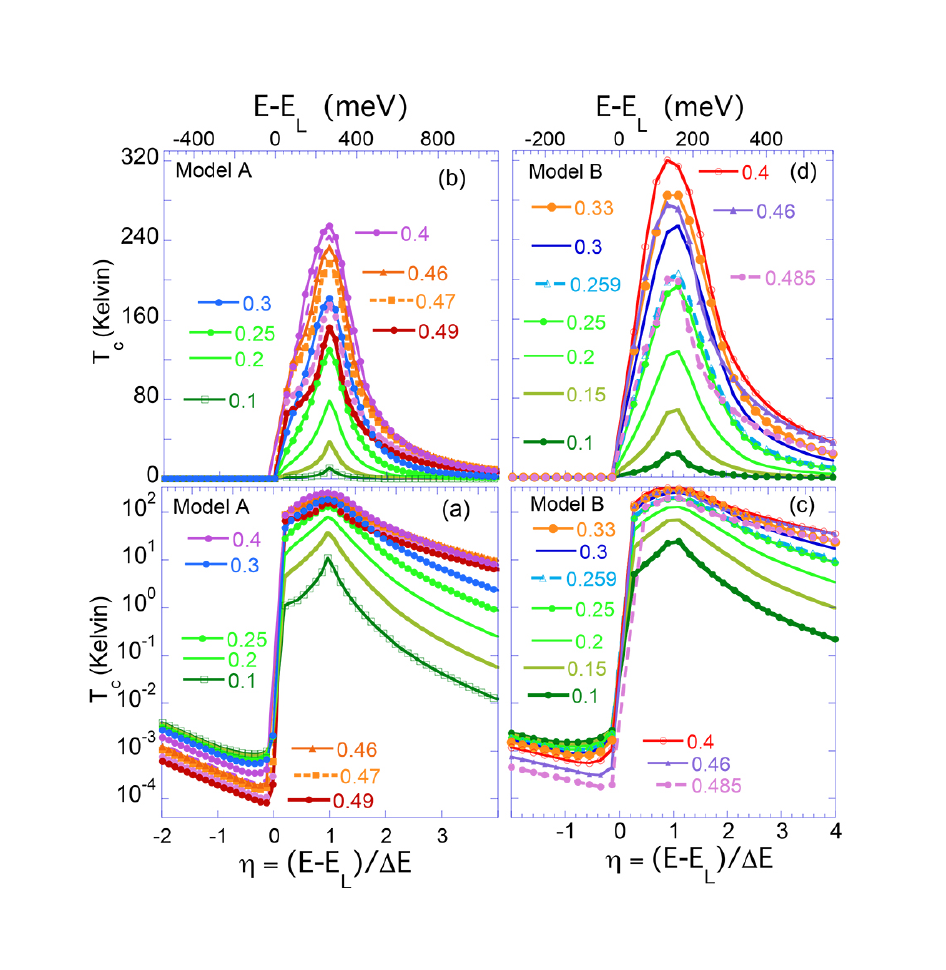}
	\caption{{Critical temperature as a function of the Lifshitz parameter (horizontal axis at the bottom) and as a function of the energy with respect to the band-edge (horizontal axis at the top) as the electron-phonon coupling varies, for two different models: case A (left panels) and case B (right panels). The curves in the bottom panels are plotted in a linear scale to show the evolution of the superconducting dome and in a semi-log scale in top panels to show the suppression of the critical temperature due to the Fano antiresonance of the left side of the dome. This plot shows that the critical temperature $T_c$ for case A reaches a maximum value of $250\ K$, while for case B it reaches a maximum value of  $330\ K$}.}
	\label{fig:5}     
\end{figure}

In Fig.\ref{fig:11} the panel A1  (B1) shows the values of the gap ratio, $2\Delta/T_c$, for the second and third subband for the A (B) case  as a function of the Lifshitz parameter $\eta$. The panel A2 (B2)  in Fig.\ref{fig:11}  shows the trend of the isotope coefficient for the A (B) case  as a function of the Lifshitz parameter $\eta$.
All these graphs were obtained at a fixed coupling value equal to $g=1/4$ for the case (A) and at $g=1/3$ for the case (B).  
At the Lifshitz transition for the appearing of a new Fermi surface, $ \eta = 0 $, the value of the gap ratio $2\Delta/T_c$ is close to the predicted value the BCS theory ($ 2 \Delta / T_c = 3.5 $)  but for  $ 0.5 < \eta < 1= 0 $ we see strong deviations from the BCS single gap prediction. In fact, $ 2 \Delta_2 / T_c $ reaches a very small value, between 0 and 1, while $ 2 \Delta_3 / T_c $ remains approximately constant at the BCS value. A similar scenario was observed in magnesium diboride [\onlinecite{innocenti2010resonant}] due to the exchange integral for pairs transfer between the second and third subbands.\\
While the BCS theory predicts that the isotope coefficient should be constant close to the value of 0.5 
we see that the isotope coefficient shows a strong maximum of the Lifshitz transition $ \eta = 0 $, and a minimum  at $ 0.5 < \eta < 1= 0 $ near the topological Lifshitz transition for opening a neck in the Fermi surface of the third subband, These theoretical predictions are in agreement with the experiments showing that the isotope coefficient shows an anomalous trend as a function of pressure in pressurized sulfur hydrides [\onlinecite{jarlborg2016breakdown}]. I

In Fig.\ref{fig:12} we show  the trend of the critical temperature as a function of the ratio between the gap in the third subband and the gap in the second subband  for the case A in panel A and for the case B in panel B. The results clearly show that the maximum critical temperature is reached with the highest anisotropy between the gaps. In fact the graph shows that the maximum of $ T_c $ is reached when the ratio $ \Delta_3 / \Delta_2$ is maximum. This figure shows clearly that room-temperature superconductivity is reached by increasing electron-phonon coupling in the a small Fermi surface spot pushing up the gap in the appearing Fermi surface due to the third subband  $\Delta_3$  while  the gap $\Delta_2$ in the second Fermi surface with large Fermi energy remains small because the electron-phonon coupling remains small. 
These results show the predicted effect of  Fano-Feshbach resonance driven by the exchange interaction between closed (strong) pairing channels in the third subband and
open (weak) pairing channels in the second subband

\section{Superconducting dome}
\begin{figure}
	\includegraphics[scale=0.9]{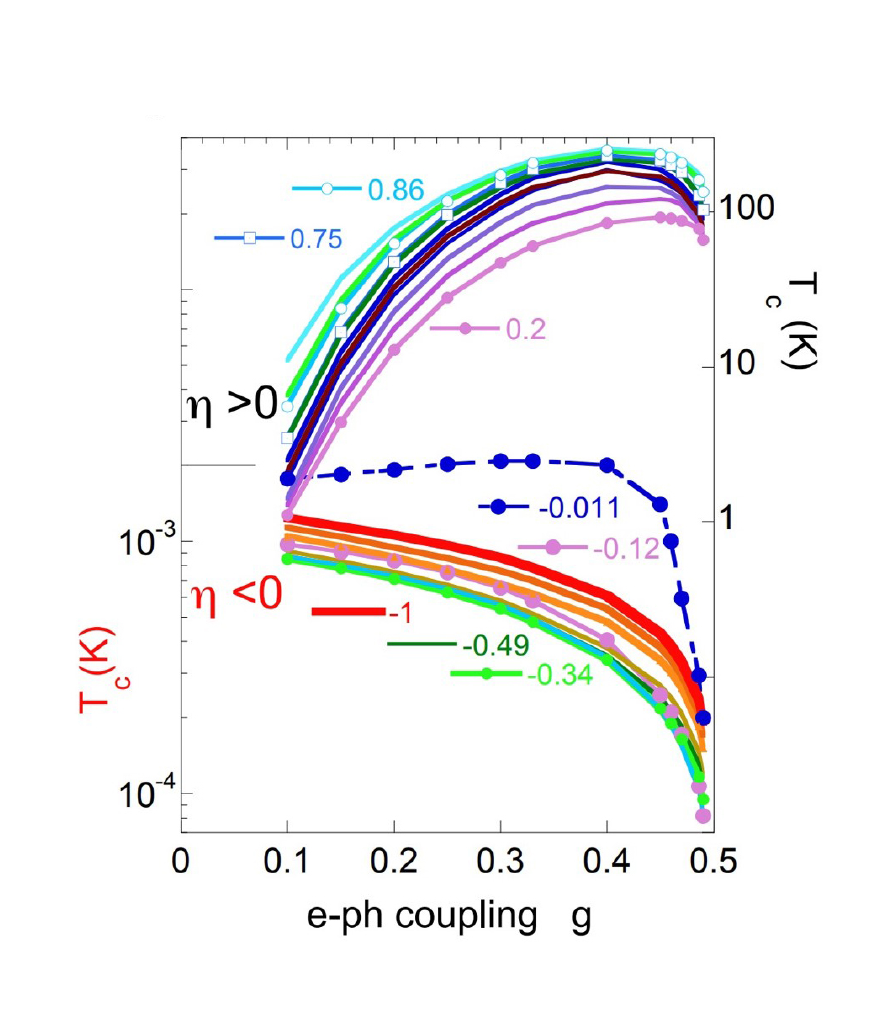}
	\caption{The critical temperature $T_c$ as a function of the electron-phonon coupling $g$ at fixed different Lifshitz parameters $\eta$ for the case A.
		The curves $T_c(g)$ in the lower part of the figure show the case with fixed  $\eta$ in the anti-resonant regime $-1<\eta<0$ where the critical temperature $decreases$ by increasing the electron phonon coupling g reaching a minimum at $\eta=-0.34$.
		The upper part of the figure shows the cases for $\eta>0$ in the resonant regime, 
		where the critical temperature $increases$ by increasing g in the range $ 0.1<g<0.4 $ with the temperature scale in the right side.}
	\label{fig:6}     
\end{figure}

In Fig.{\ref{fig:5}}  we plot the critical temperature as a function of the Lifshitz parameter in both semi-logarithmic and linear scales for variable values of the electron-phonon
(e-ph) coupling in the upper subband.  In the linear scale we see a variable superconducting dome where $T_{c \ max}$ increases with $g$ increasing up to $g=0.4$ and it decreases in the range $0.4<g<0.5$ because  the phonon softening goes to zero at  $g=0.5$.
In the case (A)  the maximum value of the critical temperature is $ 250 \ K $, therefore it explain the superconductivity in $ H_3S $. While the maximum $ T_c $ in case (B) reaches $ 330 \ K $ showing the possibility of room-temperature superconductivity. 
The plots in semi-logarithmic scale show the typical form of the Fano-Feshbach anti-resonance which becomes more relevant as $g$ increases. 
Fig.\ref{fig:6} shows the  variation of the critical temperature $T_c$ at constant $\eta$ and the variable electron-phonon coupling $g$ for the A case. In the anti-resonant regime $-1<\eta<0$ we observe a clear feature of the Fano-Feshbach resonance. In fact, at the low energy side of the Fano-Feshbach resonance between closed and open channels, the negative interference gives the observed $T_c$ minimum appearing at $\eta=-0.34$ where the critical temperature decreases with increasing e-ph coupling, on the contrary for $\eta>0$ $T_c$ increases with increasing e-ph coupling up to $g=0.4$.

\begin{figure}
	\includegraphics[scale=1.9]{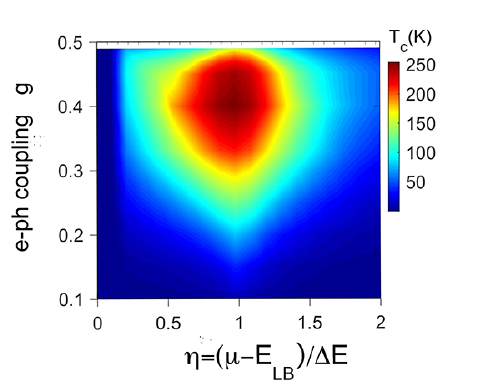}
	\caption{
		Calculated superconducting dome for $ H_3S$ simulated using the proposed A model. The critical temperature is plotted in a color plot from blue ($ T_c = 0K $) to red ($ T_c = 250K $) as a function of two variables controlling the pairing strength in the new appearing small Fermi surface above the band edge of the third upper subband: (i) the Lifshitz parameter $\eta$ measuring the normalized Fermi energy $E_{F3}$ in the range $0<\eta<2$ and (ii) the reduced Dynes [\onlinecite{dynes1972mcmillan}] electron-phonon coupling $g$=($\lambda$/(1+$\lambda))$ with lambda equal to the bare pair coupling.}
	\label{fig:7}     
\end{figure}

Fig.\ref{fig:7}  shows the critical temperature $T_c$(g,$\eta$) as a function of two variables: $(i)$ the e-ph coupling in the third subband  (g), where $g$ is the reduced Allen-Dynes electron-phonon coupling ($\lambda$/(1+$\lambda$)) [\onlinecite{dynes1972mcmillan}]  and $(ii)$ the Lifshitz energy parameter ($\eta$). 
The critical temperature  $T_c$ is calculated by the BPV approach including the superconducting shape resonance between multiple gaps. 
The maximum $ T_c$ of the dome occurs in the  $(\eta,g)$ plane at the point $(1,0.4)$ $i.e.,$  at the Lifshitz transition for neck disrupting, at $\eta$=1, which is associated with a transition of the topology of the small appearing Fermi surface from 1D at higher energy to 2D topology at lower energy.  
The \textit{universal superconductive dome} obtained in this figure is needed to understand the experimental dome observed in the  experimental curves  of the  critical pressure versus pressure $T_c$(P)  of sulfur hydrides.
In fact the external pressure induces a joint variation of both the energy position of the chemical potential with respect to the band-edge (the Lifshitz parameter $\eta$ ) as well as the electron-phonon coupling $g$ in the upper subband along a line in the ($\eta$,$g$) plane.
The variable electron-phonon coupling is associated with the softening of the phonon mode energy coupled with electrons in the upper subband
according to the Migdal relation. Therefore the experimental curve of $T_c$ \textit{vs} pressure shown in Fig.1 for a particular pressurized hydride is determined by different  cuts of the universal superconductive dome determined by the particular pathway in the ($\eta$,$g$) plane driven by variable pressure. 

\begin{figure}
	\includegraphics[scale=1.8]{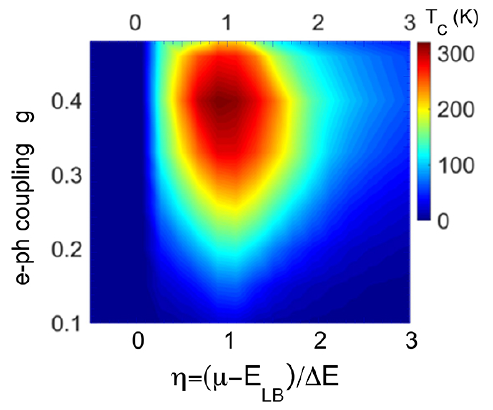}
	\caption{ 
		3D color plot of  $T_c$  as a function of two variables  ($\eta, g$)  calculated using the BPV approach for the heterostructure B, proposed here as the B model for pressurized $CSH_x$. The critical temperature increases from blue ($ T_c = 0  \ K $) to red ($ T_c =300 \ K $) in the ($\eta$,$g$) plane. The superconducting dome is due to the shape resonance between multiple superconducting gaps where  the superconducting critical temperature reaches room temperature as shown in the color plot of ($T_c $) as a function of the Lifshitz parameter $\eta$ and the electron-phonon coupling $g$ in the appearing Fermi surface at the $E_L$ energy of the Lifshitz transition
		The critical temperature reaches room temperature superconductivity where the multi-gap superconductor is close to a Lifshitz transition for neck disrupting with the Lifshitz parameter in the range $0.6<\eta<1$ and e-ph coupling close to 0.4.}
	\label{fig:9} 
\end{figure}

In order to reproduce room-temperature superconductivity in $CSH_x$ we have numerically evaluated the gaps and critical temperature for the case (B) where we have decreased the hopping between the wires to simulate the modified spacer material in $CSH_x$ in comparison with $H_3S$.
Therefore we have used the transversal dispersion $ \Delta E = 145$ meV  to simulate the superconducting dome of  $CSH_x$ . The results are shown in Fig.{\ref{fig:7}} and Fig.{\ref{fig:9}}. In the  case B  the critical temperature $T_{c \ max}$ at  the top of the $superconducting$ $dome$ reaches room-temperature superconductivity.

    Tuning the chemical potential, $\mu$, in the proximity of the band-edge, the superconducting system reaches different regimes which are distinguished by the Lifshitz parameter. At the Lifshitz transition for appearing of the new Fermi surface spot the Fermi level in the hot spot is very low and therefore the few electrons there are strongly coupled with lattice phonons showing the Khon anomaly and softening  with superconductivity competing with charge density wave (CDW) and phase separation as it has been observed in doped diborides [\onlinecite{bauer2001thermal},\onlinecite{agrestini2004substitution}] which show phonon softening at the maximum $T_c$ [\onlinecite{simonelli2009isotope}]. 

In the Lifshitz transition for the topological transition of the type opening a neck the Fano-Feshbach resonance gives the maximum $T_c$. 
In fact the BCS condensate, made of the majority of electrons in the first subband, coexists with a minority of electrons in the second subband 
forming a condensate in BCS-BEC crossover [\onlinecite{guidini2016bcs,ochi2021resonant}].
These results show that the maximum critical temperature in the multi-gap superconducting scenario can reaches room-temperature superconductivity 
driven by the exchange interaction between different condensates, neglected in the BCS approximation.

\section{Conclusions}

In conclusion, we have shown that the theory of multi-gap superconductivity in a superlattice of nanoscale stripes,
which was first proposed for high temperature superconductivity in  hole doped cuprate perovskites  [\onlinecite{perali1996gap,
valletta1997electronic,
bianconi1998superconductivity,
bianconi2006multiband,
perali2012anomalous}],
could provide a quantitative description of room temperature superconductivity in pressurized hydrides.
We have calculated the superconducting domes for two different cases of the heterostructure of quantum stripes with larger and smaller hopping between stripes
where the critical temperature is determined by both the Lifshitz parameter and the variable electron-phonon coupling in the appearing Fermi surface.
The key point of our work is the solution of  the Bogoljubov gap equations in a multi-gap system including the Fano-Feshbach resonance driven by the variable exchange interaction between condensates, which is usually neglected in the standard Migdal-Eliashberg theory.
We have shown that multiple gaps in large Fermi surfaces with high Fermi energy in the weak coupling regime can be amplified by exchange interaction with  
a large gap in the strong coupling regime in a small Fermi surface spot. 
We have presented cases where the Fano-Feshbach resonance appears by tuning the chemical potential near an electronic topological Lifshitz transition in heterostructures of quantum wires. We have presented two different heterostructures of quantum wires where the critical temperature reaches the $200 < T_c <300 K$ range. 

\begin{acknowledgments}
	We thank the staff of Department of Mathematics and Physics of Roma Tre University, the Computing Center of Institute of Microelectronics and Microsystems IMM of Italian National Research Council CNR and Supertripes-onlus for financial support of this research project.
\end{acknowledgments}	

\bibliography{Bibliography}
\end{document}